\documentclass[pra,twocolumn]{revtex4}
\usepackage{amssymb}
\usepackage{amsfonts}
\usepackage{amsmath}
\usepackage{graphicx}

\begin{document}

\title{Violation of Diagonal Non-Invasiveness: A Hallmark of Non-Classical
Memory Effects}
\author{Adri\'{a}n A. Budini}
\affiliation{Consejo Nacional de Investigaciones Cient\'{\i}ficas y T\'{e}cnicas
(CONICET), Centro At\'{o}mico Bariloche, Avenida E. Bustillo Km 9.5, (8400)
Bariloche, Argentina, and Universidad Tecnol\'{o}gica Nacional (UTN-FRC),
Fanny Newbery 111, (8400) Bariloche, Argentina}
\date{\today }

\begin{abstract}
An operational (measurement based) scheme that connects measurement
invasiveness and the presence of memory effects in open quantum systems is
defined. Its underlying theoretical basis relies on a non-invasive
measurability of (memoryless) quantum Markovian dynamics when the
corresponding observable is diagonal in the same basis as the system density
matrix. In contrast, violations of this property can be assigned to
intrinsic non-classical system memory effects. Related conditions for
violation of Leggett-Garg inequality due to quantum memory effects emerge
from this perspective. The developed approach applies to open quantum
dynamics whose time-evolution is derived from a full unitary (microscopic)
description, stochastic Hamiltonian dynamics, as well as for a broad class
of non-unitary system-environment models (bipartite Lindblad equations).
\end{abstract}

\maketitle

\section{Introduction}

Several well understood physical principles allow us to distinguish
classical from quantum realms. For example, the principle of locality, valid
in a classical regime, is violated in presence of quantum entanglement \cite%
{bell,entanglement}. \textit{Non-invasive measurability} is another
principle that distinguishes classical and quantum systems. In fact, a
quantum measurement process generally leads to an unavoidable modification
of the state of the system. Added to macroscopic realism, this feature is
the basis of Leggett-Garg inequality (LGI)~\cite{leggett,noriRep}. It
defines a constraint on the correlations that can be established between
measurements performed over a single system at two different times. A wide
range of theoretical and experimental results were analyzed and proposed in
the literature~\cite%
{mahler,huelga,korotkov,buttiker,caslav,jordan,chen,dressel,FHuelga,RMN,nori,brukner,groot,budro,emary,alt,guo,quiroga,haliwell,cunha}%
, being a topic of current interest from different perspectives~\cite%
{pagliara,laflame,vitale,reid,arani,yama}.

From a simplified point of view, it can be stated that both closed and open
quantum systems are always altered by a measurement process, leading to LGI
violation. This type of simplification contrast with strong advances made in
recent years in the classification and understanding of open quantum
dynamics~\cite{alicki,breuerbook,vega}. In particular, the association
between memory effects and time-convoluted contributions in the density
matrix evolution was mainly abandoned. Instead, memory effects (associated
to the system of interest) are defined from two complementary perspectives.
In non-operational approaches~\cite%
{BreuerReview,plenioReview,BreuerFirst,cirac,rivas,DarioSabrina,fisher,geometrical,canonicalCresser,vina}%
, memory effects are related to departures of the system time-evolution from
a (memoryless) Markovian Lindblad master equation~\cite{breuerbook,alicki}.
Alternatively, in operational approaches the system is subjected to a set of
measurement processes~\cite%
{modi,budiniCPF,pollock,pollockInfluence,bonifacio,han,ban,hefei,rio,goan,BIF}%
. Thus, non-Markovianity is defined in a standard probabilistic way~\cite%
{vanKampen} from the corresponding outcome statistics.

Given the previous advances in the characterization of open system dynamics,
it is compelling to find out if there exist any general relationship between
measurement invasiveness and the presence of memory effects. Relevant
progresses along these lines were performed previously~\cite%
{diogo,ali,shao,plenio}. However, not any clear boundary in the properties
of measurement invasiveness seems to be defined by the presence or absence
of memory effects in the system dynamics, regardless of which definition is
adopted. Therefore, a criterion of broad applicability that allows relating
measurement invasiveness with the properties of the system dynamics,
Markovian or non-Markovian, is still lacking.

The aim of this work is to establish under which conditions a clear and
general relationship between measurement invasiveness and the presence of
memory effects in open quantum systems can be established. The main
theoretical ingredient of the present approach is a diagonal
non-invasiveness (DNI) of Markovian dynamics, which applies when the
measurement observable commutates with the (pre-measurement) system density
matrix. Therefore, the observable and system state are \textit{diagonal} in
the same basis of states. Under this condition, memoryless (Markovian)
quantum dynamics behaves as classical systems (lack of measurement
invasiveness). Thus, violations of DNI are only observed in presence of
memory effects, which consistently are classified as \textit{non-classical}.
An operational scheme, based on performing three consecutive system
measurement processes, allow to witnessing these properties. Furthermore,
the developed approach enables us to establish under which conditions
violations of LGI can be related to the presence of non-classical memory
effects.

We demonstrate that the above relation between (operationally defined~\cite%
{budiniCPF}) memory effects and measurement invasiveness apply to
non-Markovian dynamics derived from unitary system-environment ($s$-$e$)
interactions and stochastic Hamiltonian models (see Appendix~\ref%
{SEInteraction}). The approach also applies to a broad class of
\textquotedblleft non-unitary $s$-$e$ coupling models\textquotedblright\
(bipartite Lindblad equations). Nevertheless, in this case it is also
possible to find a specific group of quantum dynamics, termed as
superclassical, which are capable of inducing memory effects but
consistently with a classical non-invasive structure, that is, DNI is
fulfilled in presence of memory effects. This case is characterized in
detail in an accompanying paper~\cite{super}, where in addition the
classicality notion (in presence of memory effects) that emerges from the
present approach is compared with other operational (measurement based)~\cite%
{plenio} and non-operational~\cite{horoNoMark} definitions recently
introduced in the literature.

The paper is outlined as follows. In Sec. II we introduce a quantifier of
measurement invasiveness. In Sec.~III we characterize the DNI property of
Markovian dynamics and the notion of classicality that derives from it. In
Sec. IV we characterize measurement invasiveness in non-Markovian dynamics
defined by stochastic Hamiltonians and unitary $s$-$e$ models. Non-unitary $%
s $-$e$ dynamics and superclassicality are briefly accounted. In addition,
conditions for violations of LGI due to non-classical memory effects are
established. In Sec. V we study a set of examples that corroborate the
approach. In Sec. VI we provide the conclusions. Additional supporting
demonstrations are provided in the Appendixes.

\section{Measurement invasiveness}

In operational approaches, memory effects can be witnessed with (a minimum
of) three consecutive system measurement processes~\cite{budiniCPF}.
Consequently, here measurement invasiveness is defined from the same
(operational) basis. The measurements are performed at times $0,$ $t,$ and $%
t+\tau ,$ delivering correspondingly the outcomes $\{x\},$ $\{y\},$ and $%
\{z\}.$ Their joint probability is denoted as $P_{3}(z,y,x),$ where the
subindex indicates the number of performed measurements. Disregarding the
intermediate outcomes, the marginal probability is%
\begin{equation}
P_{3}(z,x)\equiv \sum\nolimits_{y}P_{3}(z,y,x).  \label{P3ZX}
\end{equation}%
Alternatively, one can perform only two measurements at times $0$ and $%
t+\tau ,$ which defines the joint probability $P_{2}(z,x).$ For quantum
systems, measurement invasiveness implies that $P_{2}(z,x)\neq P_{3}(z,x).$
In order to quantify this disagreement, we use the probability distance~\cite%
{nielsen}%
\begin{equation}
I\equiv \sum\nolimits_{zx}|P_{3}(z,x)-P_{2}(z,x)|.  \label{Inv}
\end{equation}%
For classical systems $I=0,$ while $I>0$ is a direct witness of measurement
invasiveness. This quantum property is valid independently of the system
dynamics, closed or open, Markovian or non-Markovian. In the following
analysis, the quantifier $I$\ is studied by assuming different system
dynamics.

\section{Markovian dynamics}

In the present approach a dynamics is \textit{defined} as Markovian when the
density matrix propagator, denoted as $\Lambda _{t,t^{\prime }},$ is
completely positive (CP) and in addition is completely independent of the
(previous) measurement system\ history. Hence, in the case of two
measurement processes it follows that 
\begin{equation}
\frac{P_{2}(z,x)}{P_{1}(x)}=\mathrm{Tr}_{s}(E_{z}\Lambda _{t+\tau ,0}[\rho
_{x}]).  \label{P2Markov}
\end{equation}%
Similarly, when performing three measurements it follows%
\begin{equation}
\frac{P_{3}(z,y,x)}{P_{1}(x)}=\mathrm{Tr}_{s}(E_{z}\Lambda _{t+\tau ,t}[\rho
_{y}])\ \mathrm{Tr}_{s}(E_{y}\Lambda _{t,0}[\rho _{x}]).  \label{P3Markov}
\end{equation}%
In both cases, $P_{1}(x)=\mathrm{Tr}_{s}(E_{x}\rho _{0}),$ where $\rho _{0}$
is the initial system state. $\mathrm{Tr}(\bullet )$ is the trace operation. 
$\{E_{m}\}$ and $\{\rho _{m}\}$ $(m=x,y,z)$ are the (positive) measurement
operators and system post-measurement states respectively~\cite%
{breuerbook,nielsen}. We assume (non-degenerate) Hermitian observables.
Hence, both $\{E_{m}\}$ and $\{\rho _{m}\}$ are the projectors associated to
each observable spectral representation.

Under the assumptions considered [e.g. (arbitrary) projective measurements
and underlying $s$-$e$ propagators that fulfill a semigroup property
(Appendix~\ref{SEInteraction})], the Markovian property of the system
propagator is \textit{equivalent} to Markovianity in probability (see
Appendix~\ref{MarkovPP}), $P_{3}(z,y,x)=P_{2}(z|y)P_{2}(y|x)P_{1}(x),$ where 
$P(b|a)$ denotes the conditional probability of $b$ given $a.$ Departure
from this condition (at least for some set of consecutive measurement
processes), $P_{3}(z,y,x)\neq P_{2}(z|y)P_{2}(y|x)P_{1}(x),$ defines the
presence of \textit{memory effects}, that is, the system dynamics is
non-Markovian. We notice that this operational (measurement based)
definition of memory effects has not associated any criteria of classicality
or quantumness. This distinction is introduced over the basis of the
following properties.

\subsection*{Diagonal non-invasiveness}

From Eq.~(\ref{P2Markov}) and~(\ref{P3Markov}) it follows that in general $%
I\neq 0.$ In fact,\ measurement invasiveness occurs even when the system
dynamics is Markovian. Nevertheless, given that the measurements are
arbitrary ones, the intermediate $y$-measurement can be chosen such that $%
[\rho _{t|x},E_{y}]=0,$ where $\rho _{t|x}\equiv \Lambda _{t,0}[\rho _{x}].$
Thus, $\{\mathrm{Tr}_{s}(E_{y}\Lambda _{t,0}[\rho _{x}])\}$ can be read as
the eigenvalues of $\rho _{t|x}$ implying $\sum\nolimits_{y}\rho _{y}\mathrm{%
Tr}_{s}(E_{y}\Lambda _{t,0}[\rho _{x}])=\rho _{t|x}.$ Consequently,%
\begin{equation}
I_{D}\overset{M}{=}0.  \label{DNI}
\end{equation}%
\textit{This property defines the DNI of Markovian dynamics}: a
(non-selective) measurement process at a given time is non-invasive if the
corresponding observable commutates with the pre-measurement system density
matrix~\cite{demolition}. In the three measurement scheme DNI at (any) time $%
t$ is valid for \textit{arbitrary} $x$- and $z$-measurements. These are
central results for the developed scheme.

The DNI property of Markovian dynamics implies that, at the intermediate
stage, the quantum system behaves as a classical system, that is, it is
unaffected by the intermediate measurement (invasiveness is always avoided
independently of which are the previous and posterior quantum measurement
processes). This property gives us the basis for classifying system memory
effects as quantum or classical. Memory effects are \textit{non-classical}
(quantum) when in the three measurement scheme there exist (at least a first
and last) measurement processes such that DNI is not achieved. In fact,
violation of DNI cannot be achieved by any classical system. In this way,
measurement invasiveness and the non-classical character of the memory
effects becomes intrinsically linked in the present approach. The
complementary notion of classicality that follows from this definition as
well as those introduced in Refs.~\cite{plenio,horoNoMark} are studied and
compared in detail in Ref.~\cite{super}.

\section{Measurement invasiveness in non-Markovian dynamics}

We remark that the previous criteria of non-classicality, as well as the
definition of memory effects, rely on the joint outcome probabilities. They
do not depend on the nature (classical or quantum) of the environment that
is coupled to the system of interest. Below we consider different $s$-$e$
models where the underlying time-evolution fulfills a semigroup property
that in turn is not affected by any measurement process performed over the
bipartite\ arrangement (Appendix~\ref{SEInteraction}). For almost all these
dynamics DNI is violated in presence of system memory effects.

\subsection{Stochastic Hamiltonian models}

Here the open system is driven by random fluctuations such that $\Lambda
_{t,t^{\prime }}=\overline{\Lambda _{t,t^{\prime }}^{st}},$ where $\Lambda
_{t,t^{\prime }}^{st}$ is a CP stochastic propagator associated to each
noise realization. The overline denotes the corresponding average. In
similarity with Eq.~(\ref{P2Markov}) it follows%
\begin{equation}
\frac{P_{2}(z,x)}{P_{1}(x)}=\mathrm{Tr}_{s}\overline{(E_{z}\Lambda _{t+\tau
,0}^{st}[\rho _{x}])},
\end{equation}%
while in similarity with Eq.~(\ref{P3Markov}),%
\begin{equation}
\frac{P_{3}(z,y,x)}{P_{1}(x)}=\mathrm{Tr}_{s}\overline{(E_{z}\Lambda
_{t+\tau ,t}^{st}[\rho _{y}])\ \mathrm{Tr}_{s}(E_{y}\Lambda _{t,0}^{st}[\rho
_{x}])}.  \label{P3Stochastic}
\end{equation}%
Hence, DNI is not valid in general. It is recovered when the average
defining $P_{3}(z,y,x)$ split in two terms as in Eq.~(\ref{P3Markov}). This
property is valid for white noise fluctuations~\cite{vanKampen}, that is,
when the dynamics is Markovian~\cite{budiniCPF}. In addition, using the
arbitrariness of the first and last measurement processes, it is possible to
prove that this is a necessary condition (see Appendix~\ref{DNIConditions}).
Thus, DNI is valid if and only if the dynamics is Markovian.

\subsection{Unitary system-environment dynamics}

The system and its environment evolve according to a CP unitary bipartite
propagator $\mathcal{G}_{t,t^{\prime }},$ which in turn is the same between
measurement processes. For uncorrelated initial conditions, it follows%
\begin{equation}
\frac{P_{2}(z,x)}{P_{1}(x)}=\mathrm{Tr}_{se}(E_{z}\mathcal{G}_{t+\tau
,0}[\rho _{x}\otimes \sigma _{0}]),  \label{DosCP}
\end{equation}%
where $\sigma _{0}$ is the environment initial state. Furthermore,%
\begin{equation}
\frac{P_{3}(z,y,x)}{P_{1}(x)}=\mathrm{Tr}_{se}(E_{z}\mathcal{G}_{t+\tau
,t}[\rho _{y}\otimes \mathrm{Tr}_{s}(E_{y}\mathcal{G}_{t,0}[\rho _{x}\otimes
\sigma _{0}])]).  \label{TresCP}
\end{equation}%
From these expressions it follows that Markovianity is not fulfilled in
general~\cite{budiniCPF}, $P_{3}(z,y,x)\neq P_{3}(z|y)P_{2}(y|x)P_{1}(x).$
Furthermore, taking into account Eq. (\ref{Inv}), we realize that DNI ($I=0$
for arbitrary first and third measurements) can only be satisfied after
demanding some property to the bipartite propagator.

It is simple to check that DNI and Markovian probabilities are
simultaneously recovered when the bipartite propagator satisfies $\mathrm{Tr}%
_{e}(\mathcal{G}_{t+\tau ,t}[\rho \otimes \sigma ])=\Lambda _{t+\tau
,t}[\rho ]\ \mathrm{Tr}_{e}[\sigma ],$ where the system propagator $\Lambda
_{t,t^{\prime }}$ is independent of the (post-measurement) bath state $%
\sigma .$ This condition is fulfilled when a Born-Markov approximation~\cite%
{breuerbook} applies (even in presence of intermediate measurement process), 
$\mathcal{G}_{t,0}[\rho \otimes \sigma ]\approx \rho _{t}\otimes \sigma .$
Similarly to the previous stochastic dynamics, assuming the validity of DNI
for arbitrary first and last measurement processes, Markovianity becomes a
necessary condition~(see Appendix~\ref{DNIConditions}). Equivalently,\textit{%
\ }for unitary $s$-$e$ models, if for some (three consecutive) measurement
processes the outcome statistics is non-Markovian there always exist $z$-
and $x$-measurement processes such that DNI is not fulfilled. Therefore,
underlying Hamiltonian microscopic dynamics lead to non-classical (system)
memory effects.

\subsection{Bipartite Lindblad models and superclassicality}

Eqs.~(\ref{DosCP}) and~(\ref{TresCP}) remain valid even when the underling $%
s $-$e$ propagator (between system measurements) can be approximated by a
bipartite Lindblad evolution~\cite{alicki,breuerbook}. In general\ DNI is
not fulfilled $(I\neq 0).$ Hence, the memory effects developed by the
subpart that constitutes the system of interest are read as non-classical.
Nevertheless, for this kind of models it is also possible to define a
particular class of non-Markovian dynamics that always obey DNI $(I=0).$
These quantum \textit{non-Markovian} dynamics are termed as \textit{%
superclassical}. In fact, the system is unaffected by the intermediate
measurement even when the previous and posterior (post-measurement) quantum
states are \textit{arbitrary} ones. While these features are shared by
Markovian dynamics, here they occur in presence of memory effects, that is,
the joint probabilities for measurement outcomes depart from a Markovian
structure $[P_{3}(z,y,x)\neq P_{3}(z|y)P_{2}(y|x)P_{1}(x)].$

In Appendix~\ref{DNIConditions} we find that superclassicality emerges when
the system propagator fulfills%
\begin{equation}
\Lambda _{t+\tau ,0}[\rho ]\overset{DNI}{=}\sum_{c}\mathrm{Tr}_{e}(\mathcal{G%
}_{t+\tau ,t}[\Pi _{t}^{c}\otimes \mathrm{Tr}_{s}(\Pi _{t}^{c}\mathcal{G}%
_{t,0}[\rho \otimes \sigma _{0}])]).  \label{Super}
\end{equation}%
Here, $\rho $ denotes an\textit{\ }arbitrary input state, while the
projectors $\{\Pi _{t}^{c}\}$ are defined by $\rho _{t}=\Lambda _{t,0}[\rho
]=\sum_{c}\Pi _{t}^{c}p_{t}^{c}.$ While it is not a necessary condition, the
propagator constraint~(\ref{Super}) is (mainly) fulfilled when the $s$-$e$
interaction do not generate quantum discord~\cite%
{ZureckDiscord,henderson,modiCorrelator} \textit{even} when considering 
\textit{arbitrary} system initial conditions $\rho .$ This property supports
denoting these dynamics as superclassical. Furthermore, they fulfill
alternative definitions of classicality based on fixed measurement basis~%
\cite{plenio} in arbitrary ones. On the other hand, the arbitrariness of the
previous and posterior measurement processes restrict superclassicality to a
subclass of (system) depolarizing dynamics. These properties are also
studied and characterized in detail in Ref.~\cite{super}.

\subsection{Leggett-Garg inequality}

By denoting with $\{O_{m}\}$ the observable values\ associated to each
outcome, $m=x,y,z,$ LGI reads~\cite{noriRep}%
\begin{equation}
-3\leq \langle O_{y}O_{x}\rangle +\langle O_{z}O_{y}\rangle -\langle
O_{z}O_{x}\rangle \leq 1.  \label{LGI}
\end{equation}%
This result is valid for dichotomic observables, $O_{m}=\pm 1.$ The
correlators, which are obtained by performing only two measurements, are $%
\langle O_{j}O_{k}\rangle \equiv \sum_{jk}O_{j}O_{k}P_{2}(j,k).$ Eq.~(\ref%
{LGI}) can be derived from $P_{2}(y,x)=\sum_{z}P_{3}(z,y,x),$ and by
assuming that $P_{2}(z,y)\overset{LGI}{=}\sum_{x}P_{3}(z,y,x),$ and $%
P_{2}(z,x)\overset{LGI}{=}\sum_{y}P_{3}(z,y,x)$~\cite{noriRep}. Due to
measurement invasiveness and independently of the system dynamics (Markovian
or non-Markovian), these equalities are not valid in general. Nevertheless,
consistently with the previous analysis [DNI,\ Eq.~(\ref{DNI})] it follows
that Markovian dynamics always obey LGI if at each pair of measurement
processes the observables commutate with the pre-measurement system density
matrix. Thus,\ \textit{if the three observables commutates with the
pre-measurement state violations of LGI can be related to non-classical
memory effects}. On the other hand, even under this election of measurement
processes, non-Markovian dynamics may or may not obey LGI. In this sense,
the distance $I$ [Eq.~(\ref{Inv})] provides a deeper description of
measurement invasiveness.

\section{Examples}

Here we study different $s$-$e$ interactions that lead to a dephasing
non-Markovian dynamics. The system is a qubit (two-level system) with basis
of states $|\pm \rangle .$ Its density matrix reads%
\begin{equation}
\rho _{t}=\left( 
\begin{array}{cc}
\langle +|\rho _{0}|+\rangle & d(t)\langle +|\rho _{0}|-\rangle \\ 
d(t)^{\ast }\langle -|\rho _{0}|+\rangle & \langle -|\rho _{0}|-\rangle%
\end{array}%
\right) .  \label{RhoS}
\end{equation}%
The populations remain constant while the coherences are characterized by
the decay function $d(t).$

In correspondence with the analyzed dynamics, the first evolution is set by a%
\textit{\ stochastic Hamiltonian}~\cite{anderson,GaussianNoise}%
\begin{equation}
\frac{d\rho _{t}^{st}}{dt}=-i\xi (t)[\sigma _{z},\rho _{t}^{st}].
\label{HEsto}
\end{equation}%
Here, $\sigma _{z}$ is the $z$-Pauli matrix. The noise has a null average $%
\overline{\xi (t)}=0$ and correlation $\overline{\xi (t)\xi (t^{\prime })}%
=(\gamma /2\tau _{c})\exp [-(t-t^{\prime })/\tau _{c}].$ From the system
density matrix $\rho _{t}=\overline{\rho _{t}^{st}},$ the coherence decay
reads $d(t)=\exp \left[ -2\gamma (t-\tau _{c}(1-e^{-t/\tau _{c}})\right] .$
In the limit of a vanishing correlation time (white noise), $\tau
_{c}/\gamma \rightarrow 0,$ an exponential decay is recovered.

A \textit{unitary} $s$-$e$\textit{\ model} is defined by a spin bath~\cite%
{ZurekRMP,Paz},%
\begin{equation}
\frac{d\rho _{t}^{se}}{dt}=-ig\Big{[}\sigma _{z}\otimes
\sum\nolimits_{j=1}^{n}\sigma _{z}^{(j)},\rho _{t}^{se}\Big{]}.
\label{SpinBath}
\end{equation}%
Here, $g$ is a coupling parameter while $\sigma _{z}^{(j)}$ is the $z$-Pauli
matrix of the $j$ environment spin. The system density matrix $\rho _{t}=%
\mathrm{Tr}_{e}[\rho _{t}^{se}]$ follows by tracing the bipartite $s$-$e$
state. Assuming that each bath spin begins in an identity state, the system
coherence decay reads $d(t)=[\cos (2gt)]^{n}.$

A \textit{dissipative} \textit{(Lindblad)} $s$-$e$\textit{\ model} is set by
a non-diagonal multipartite Lindblad equation~\cite{Clerck,noDiag}%
\begin{equation}
\frac{d\rho _{t}^{se}}{dt}=\sum_{j,k=1}^{n}\Gamma _{jk}(\sigma
_{z}^{(j)}\rho _{t}^{se}\sigma _{z}^{(k)}-\frac{1}{2}\{\sigma
_{z}^{(k)}\sigma _{z}^{(j)},\rho _{t}^{se}\}_{+}),  \label{NonDiagol}
\end{equation}%
where $\Gamma _{jk}=(\gamma -\chi )\delta _{jk}+\chi (i)^{j-1}(-i)^{k-1}.$
When $\chi =0,$ each qubit obeys a Markovian dephasing evolution with rate $%
\gamma .$ When $\chi \neq 0$ all subsystems are coupled to each other,
leading to the development of memory effects. Taking the first qubit as the
system of interest, the coherence decay reads $d(t)=e^{-2\gamma t}[\cos
(2\chi t)]^{\bar{n}},$ where $\bar{n}=\mathrm{Int}(n/2)$ is the integer part
of $n/2.$ This result relies on assuming that each environment qubit
subsystem begins in a completely mixed state~\cite{noDiag}.

For the three previous models it is possible to calculate $P_{3}(z,y,x)$ in
an exact analytical way, where $z=\pm 1,$ $y=\pm 1,$ and $x=\pm 1.$ We
assume that three measurement processes are performed successively in the
Bloch directions $\hat{x}-\hat{n}-\hat{x},$ where the vector $\hat{n}=\hat{n}%
(\theta ,\phi )$ is defined by polar angles $(\theta ,\phi ).$ Using that $%
d(t)=d(t)^{\ast },$ we get%
\begin{equation}
\frac{P_{3}(z,y,x)}{P_{1}(x)}=\frac{1}{4}[1+yxf(t)+zyf(\tau )+zxf(t,\tau )],
\label{P3Models}
\end{equation}%
where $f(t)=\sin (\theta )\cos (\phi )d(t),$ while%
\begin{equation}
f(t,\tau )=\frac{1}{2}\sin ^{2}(\theta )[d(t+\tau )+\cos (2\phi )d(t,\tau )].
\end{equation}%
The function $d(t)$ is the coherence decay of each model, while $d(t,\tau )$
differ in each case~\cite{supleEQ}. In all cases $P_{3}(z,y,x)$ does not
fulfill (in general) a Markovian property. 
\begin{figure}[tbp]
\includegraphics[bb=40 585 725 1080,
angle=0,width=8.5cm]{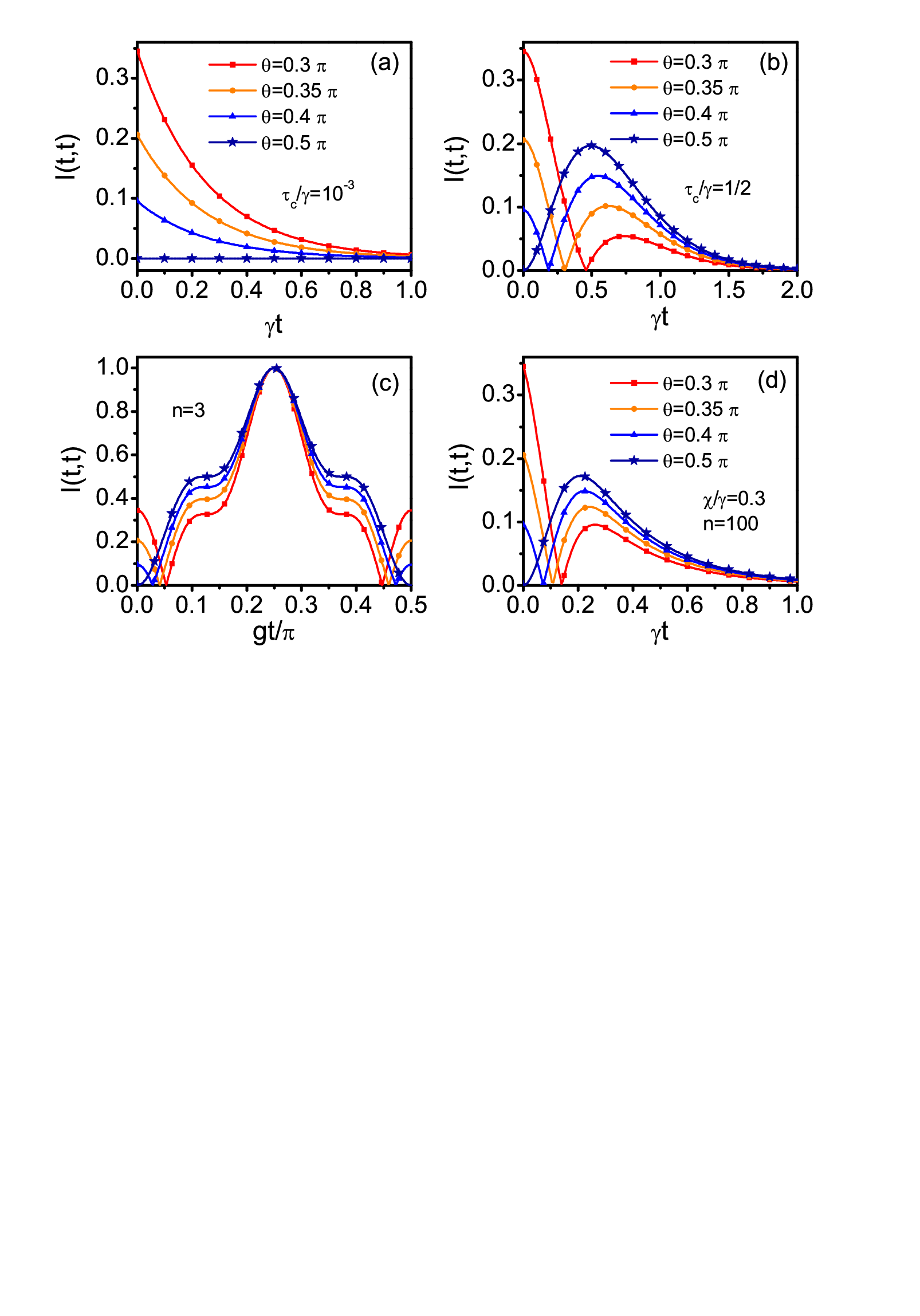}
\caption{Invasivity measure $I(t,\protect\tau )$ at equal time-intervals $(%
\protect\tau =t)$ for different underlying open system dynamics, where the
measurements are performed in Bloch directions $\hat{x}-\hat{n}(\protect%
\theta ,\protect\phi )-\hat{x}$ with $\protect\phi =0.$ (a) and (b)
Stochastic Hamiltonian model [Eq.~(\protect\ref{HEsto})]. (c) Hamiltonian $s$%
-$e$ model [Eq.~(\protect\ref{SpinBath})]. (d) Dissipative model [Eq.~(%
\protect\ref{NonDiagol})]. The parameters are indicated in each plot. The
DNI Bloch-direction is $\protect\theta =\protect\pi /2,$ $\protect\phi =0.$}
\end{figure}

From Eqs.~(\ref{P3ZX}) and~(\ref{P3Models}) it is easy to obtain $%
P_{3}(z,x)=[1+zxf(t,\tau )]P_{1}(x)/2.$ Given that the last measurement is
performed in the $\hat{x}$-Bloch direction, $P_{2}(z,x)$ can also be
obtained from Eq.~(\ref{P3Models}) under the steps $\sum_{z},$ the
subsequent replacements $y\rightarrow z,$ $t\rightarrow t+\tau ,$ and taking 
$\theta =\pi /2,$ $\phi =0,$ which deliver $P_{2}(z,x)=[1+zxd(t+\tau
)]P_{1}(x)/2.$ The invasiveness distance Eq.~(\ref{Inv}) therefore becomes $%
I=(1/2)\sum_{z,x}|zx||f(t,\tau )-d(t+\tau )|P_{1}(x).$ Using that $%
\sum_{z=\pm 1}|z|=2$ and $\sum_{x=\pm 1}|x|P_{1}(x)=1,$ we get%
\begin{equation}
I=I(t,\tau )=|f(t,\tau )-d(t+\tau )|.
\end{equation}%
This expression for $I$ is valid for an arbitrary intermediate measurement
defined by the angles $(\theta ,\phi ).$ The DNI of Markovian dynamics [Eq.~(%
\ref{DNI})] is valid when this measurement is performed in the same basis
where $\rho _{t|x}$ ($\rho _{0|x}=\rho _{x}$) is diagonal. Given that the
first measurement is performed in the $\hat{x}$-Bloch direction, Eq.~(\ref%
{RhoS}) implies $\langle \pm |\rho _{t|x}|\pm \rangle =1/2$ and $\langle \pm
|\rho _{t|x}|\mp \rangle =xd(t)/2.$ Defining $M_{\hat{\imath}}\equiv \mathrm{%
Tr}_{s}[\sigma _{i}\rho _{t|x}]$ where $\sigma _{i}$ are the Pauli matrixes,
we get $M_{\hat{x}}=xd(t),$ while $M_{\hat{y}}=M_{\hat{z}}=0.$ Hence, $\rho
_{t|x}$ is diagonal ($\forall t$) in the $\hat{x}$-Bloch direction~\cite%
{blum}. Consequently, the intermediate observable commutates with $\rho
_{t|x}$ when $\theta =\pi /2$ and $\phi =0.$ In general, these angles
becomes time-dependent when the first observable is not in the $\hat{x}$-$%
\hat{y}$ Bloch plane.

In Fig.~1 we plot $I(t,\tau )$ for the different open dynamics. The
intermediate measurement is in the~$\hat{z}$-$\hat{x}$ Bloch plane: $\phi
=0. $ Fig.~1(a) and (b) correspond to the stochastic Hamiltonian evolution\
[Eq.~(\ref{HEsto})]. In (a) the parameters approach a Markovian white noise
limit, $\tau _{c}/\gamma \simeq 0\Rightarrow d(t)\simeq \exp (-2\gamma
t)\Rightarrow P_{3}(z,y,x)\simeq P_{2}(z|y)P_{2}(y|x)P_{1}(x)$ [Eq.~(\ref%
{P3Models})]. Invasiveness is clearly observed, $I(t,\tau )\neq 0.$
Nevertheless, when $\theta \rightarrow \pi /2,$ \textit{the DNI of Markovian
dynamics is corroborated} $I(t,\tau )$ $\rightarrow 0.$ In contrast, in (b)
when $\tau _{c}/\gamma >0,$ even when the intermediate observable commutates
with the system density matrix $(\theta =\pi /2$ and $\phi =0),$
consistently with our results, due to the presence of memory effects $%
I(t,\tau )$ does not vanish.

In Fig.~1(c) we plot\ $I(t,\tau )$ for the spin environment model [Eq.~(\ref%
{SpinBath})]. Given that the number of spins\ is finite all behaviors are
periodic in time. Again, even when the measurement and system state
commutate $(\theta =\pi /2$ and $\phi =0),$ DNI is violated, $I(t,\tau )>0.$
The same result is valid for the dissipative model [Eq.~(\ref{NonDiagol})]
when $\chi /\gamma \neq 0,$ Fig.~1(d). When $\chi /\gamma \rightarrow 0,$ a
Markovian regime is approached. The behavior of $I(t,\tau )$ becomes
indistinguishable from that shown in Fig.~1(a), corroborating the DNI of
Markovian dynamics in this alternative model. 
\begin{figure}[tbp]
\includegraphics[bb=40 585 725 1080,
angle=0,width=8.5cm]{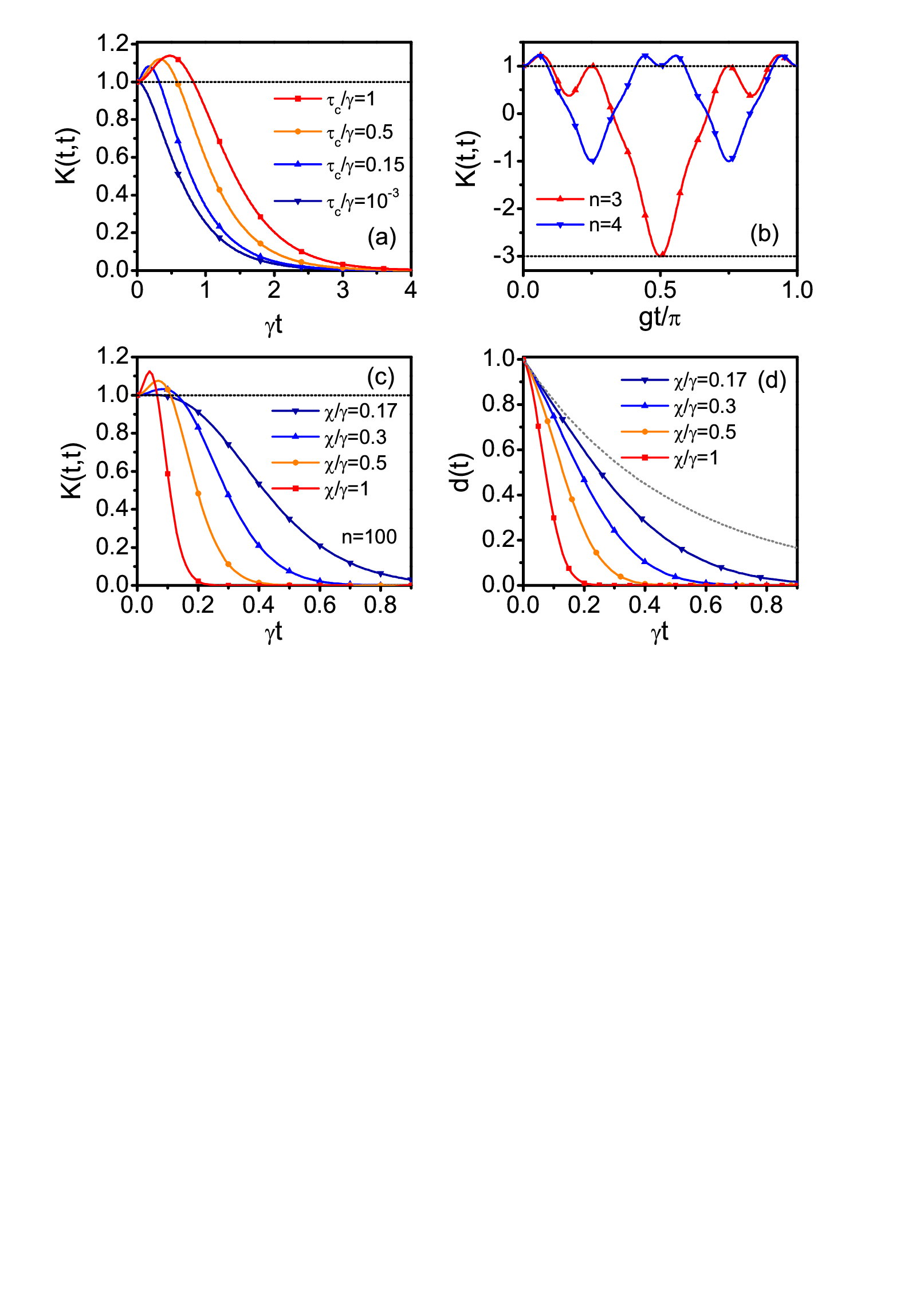}
\caption{LGI parameter $K(t,\protect\tau )\equiv d(t)+d(\protect\tau )-d(t+%
\protect\tau )$ [Eq.~(\protect\ref{LGIDecay})] for equal time-intervals. (a)
Stochastic Hamiltonian model [Eq.~(\protect\ref{HEsto})]. (b) Hamiltonian $s$%
-$e$ model [Eq.~(\protect\ref{SpinBath})]. (c) Dissipative model [Eq.~(%
\protect\ref{NonDiagol})], while (d) shows the associated system coherence
decay. The dotted line corresponds to the Markovian limit $d(t)=\exp [-2%
\protect\gamma t].$}
\end{figure}

Assuming that $\rho _{0}$ is diagonal in the $\hat{x}$-Bloch direction, the
density matrix [Eq.~(\ref{RhoS})] remains diagonal in that base. Hence,
violation of LGI due to memory effects can be checked by choosing the three
observables $\{O_{m}\}$ $(m=x,y,z)$ as diagonal in the same $\hat{x}$%
-direction. Given that only two measurements are explicitly performed, for
the three studied $s$-$e$ models, Eq.~(\ref{LGI}) $(O_{m}=m)$ can be
expressed in terms of the corresponding coherence decay,%
\begin{equation}
-3\leq d(t)+d(\tau )-d(t+\tau )\leq 1.  \label{LGIDecay}
\end{equation}

In Fig.~2 we study the validity of this inequality. For the stochastic
Hamiltonian model (a), LGI is only valid when a Markovian white noise limit
is attained, that is $\tau _{c}/\gamma =0.$ For the unitary $s$-$e$ model
(b), given the absence of a Markovian limit (in probability), LGI is
violated independently of the number of environment spins. The dissipative
model (c) presents a \textit{memory induced transition. }In fact, for $\chi
/\gamma \lesssim 0.17$ LGI is valid, which includes the Markovian case $\chi
/\gamma =0.$ Nevertheless, LGI is violated for $\chi /\gamma \gtrsim 0.17.$
In (d) we show the corresponding system coherence decay $d(t).$ All of them
are quasi-monotonic. Thus, the memory induced transition does not relies on
any revival in the coherence decay.

\section{Conclusions}

A deep relation between measurement invasiveness and the presence of memory
effects in open quantum systems has been found. The established relation is
always valid for stochastic Hamiltonian models, unitary $s$-$e$ dynamics,
and all (non-unitary) bipartite Lindblad models which do not fulfill
superclassicality~\cite{super}. The relation was established over the basis
of an operational (measurement based) scheme. It relies on performing three
consecutive system measurement processes, where the intermediate observable
must to commutates with the pre-measurement state. Using the arbitrariness
of the first and last measurement processes, we concluded that all of these
non-Markovian dynamics are intrinsically modified by a measurement process
even when the corresponding observable commutates with the system state.
This violation of DNI due to memory effects disappears when the dynamics
approaches a memoryless (operationally defined) Markovian regime.

Violations of DNI determine the non-classical character of the memory
effects, while the validity of DNI for arbitrary first and last measurement
processes define superclassicality. In stretched relation, Markovian
dynamics always obeys LGI if the pairs of measurement processes (involved in
its definition) commutate with the pre-measurement system state. Under these
conditions, violations of LGI can also be attributed to the presence of
non-classical memory effects. Nevertheless, the operational approach based
on performing three measurement processes provides a deeper discernment of
this characteristic.

The studied models support the main conclusions. The examples also allowed
us to demonstrate that memory effects can drive a transition in the validity
of LGI. Most of studied dynamics could be implemented in different
experimental photonic platforms. For example, in the schemes of Refs.~\cite%
{hefei,rio} the intermediate measurement-basis associated to violation of
DNI could be determined from standard tomographic techniques.

\section*{Acknowledgments}

Valuable discussions with Simon Milz and Susana F. Huelga are gratefully
acknowledged. This paper was supported by Consejo Nacional de
Investigaciones Cient\'{\i}ficas y T\'{e}cnicas (CONICET), Argentina.

\appendix

\section{Underlying system-environment dynamics\label{SEInteraction}}

The developed results rely on assuming that the underlying $s$-$e$
time-evolution fulfills a semigroup property, which in turn is not affected
by any measurement process performed over the bipartite arrangement.

In the case of \textit{stochastic Hamiltonian models}, the underlying
propagator $\Lambda _{t,t^{\prime }}^{st},$ for each noise realization,
fulfills%
\begin{equation}
\Lambda _{t+\tau ,t^{\prime }}^{st}=\Lambda _{t+\tau ,t}^{st}\Lambda
_{t,t^{\prime }}^{st},  \label{HEstocastico}
\end{equation}%
where $t+\tau \geq t\geq t^{\prime }.$ Specifically, it is possible to write 
$\Lambda _{t+\tau ,t}^{st}[\bullet ]=\exp \{-i\int_{t}^{t+\tau
}[H_{t^{\prime }}^{st},\bullet ]dt^{\prime }\}$ where $H_{t^{\prime }}^{st}$
is the stochastic Hamiltonian. The system propagator corresponds to the
average dynamics, $\Lambda _{t,0}[\rho ]=\overline{\Lambda _{t,0}^{st}}[\rho
].$

Alternatively, we consider \textit{completely positive bipartite }$s$-$e$ 
\textit{dynamics} where the the underlying bipartite propagator fulfills%
\begin{equation}
\mathcal{G}_{t+\tau ,t^{\prime }}=\mathcal{G}_{t+\tau ,t}\mathcal{G}%
_{t,t^{\prime }}.  \label{Gbipartito}
\end{equation}%
In this case, it is possible to write explicitly $\mathcal{G}_{t+\tau
,t}=\exp \int_{t}^{t+\tau }\mathcal{L}_{t^{\prime }}dt^{\prime }.$ For 
\textit{unitary dynamics} the Liouville superoperator reads $\mathcal{L}%
_{t^{\prime }}[\bullet ]=-i[H_{T},\bullet ],$\ where $H_{T}$ is the total $s$%
-$e$ Hamiltonian~\cite{breuerbook}. Alternatively, a \textit{non-unitary
dynamics }can be taken into account through a bipartite Lindblad equation, $%
\mathcal{L}_{t^{\prime }}[\bullet ]=-i[H_{se},\bullet ]+\sum_{\alpha }\gamma
_{\alpha }(V_{\alpha }\bullet V_{\alpha }^{\dagger }-(1/2)\{V_{\alpha
}^{\dagger }V_{\alpha },\bullet \}_{+}).$ In both cases, unitary and
non-unitary, the underlying generators could depend on time. The system
propagator reads $\Lambda _{t,0}[\rho ]=\mathrm{Tr}_{e}(\mathcal{G}%
_{t,0}[\rho \otimes \sigma _{0}]).$

\section{Equivalent definitions of Markovianity\label{MarkovPP}}

The Markovian property of quantum dynamics can be defined from the system
propagator or complementarily from the outcome probabilities. Here, we show
that both definitions are equivalent when considering the $s$-$e$ dynamics
considered in the manuscript [Eqs.~(\ref{HEstocastico}) and~(\ref{Gbipartito}%
)].

\textit{Markovianity in propagator} ($\Lambda $-Markovianity) is defined by
the property%
\begin{equation}
\Lambda _{t+\tau ,t}[\rho _{\{t,0\}}]\overset{\Lambda }{=}\Lambda _{t+\tau
,t}[\rho _{t}],  \label{PropaMarkov}
\end{equation}%
which implies that the (system) propagator $\Lambda $ is independent of the
previous system history, here denoted as $\rho _{\{t,0\}}.$ When the history
does not involve any measurement, $\rho _{\{t,0\}}=\Lambda _{t,0}[\rho
_{0}], $ the divisibility of the dynamics follows straightforwardly, $%
\Lambda _{t+\tau ,0}=\Lambda _{t+\tau ,t}\Lambda _{t,0}.$ In general, the
inverse implication is not true~\cite{QRT}. On the other hand, \textit{%
Markovianity in probability} ($p$-Markovianity) is defined by the property$\ 
$%
\begin{equation}
P_{3}(z,y,x)\overset{p}{=}P_{2}(z|y)P_{2}(y|x)P_{1}(x),
\label{p-Markovianity}
\end{equation}%
which in turn must be valid when considering \textit{arbitrary} (projective)
measurement processes performed at each stage.

It is straightforward to prove that [see Eq.~(\ref{P3Markov})]%
\begin{equation}
\Lambda \text{-Markovianity}\Rightarrow p\text{-Markovianity.}
\end{equation}%
Furthermore, the inverse implication%
\begin{equation}
\Lambda \text{-Markovianity}\Leftarrow p\text{-Markovianity,}
\end{equation}%
is also valid. The explicit demonstration of this last result is developed
below.

From Bayes rule it is always possible to write%
\begin{equation}
P_{3}(z,y,x)=P_{3}(z|y,x)P_{2}(y|x)P_{1}(x).
\end{equation}%
Consequently, $p$-Markovianity [Eq.~(\ref{p-Markovianity})] implies that%
\begin{equation}
P_{3}(z|y,x)\overset{p}{=}P_{2}(z|y).  \label{PMarkovian}
\end{equation}%
Using this condition if follows that $\Lambda $-Markovianity must be
fulfilled. The explicit demonstration of this affirmation is performed in a
separate way for stochastic Hamiltonian models and CP bipartite $s$-$e$
dynamics.

\subsection{Stochastic Hamiltonian models}

Using that $P_{3}(z|y,x)=P_{3}(z,y,x)/P_{2}(y,x)=P_{3}(z,y|x)/P_{2}(y|x),$
it is possible to write [see Eq.~(\ref{P3Stochastic})]%
\begin{equation}
P_{3}(z|y,x)=\frac{\mathrm{Tr}_{s}\overline{(E_{z}\Lambda _{t+\tau
,t}^{st}[\rho _{y}])\ \mathrm{Tr}_{s}(E_{y}\Lambda _{t,0}^{st}[\rho _{x}])}}{%
\mathrm{Tr}_{s}\overline{(E_{y}\Lambda _{t,0}^{st}[\rho _{x}])}}.
\end{equation}%
On the other hand, the conditional probability $%
P_{2}(z|y)=P_{2}(z,y)/P_{1}(y)$ appearing in Eq.~(\ref{p-Markovianity}) reads%
\begin{equation}
P_{2}(z|y)=\frac{\mathrm{Tr}_{s}\overline{(E_{z}\Lambda _{t+\tau
,t}^{st}[\rho _{y}])\ \mathrm{Tr}_{s}(E_{y}\Lambda _{t,0}^{st}[\rho _{0}])}}{%
\mathrm{Tr}_{s}\overline{(E_{y}\Lambda _{t,0}^{st}[\rho _{0}])}}.
\end{equation}%
Imposing the condition~(\ref{PMarkovian}), using the arbitrariness of the $z$%
-measurement process, it follows that%
\begin{equation}
\overline{\Lambda _{t+\tau ,t}^{st}[\rho _{\{y,x\}}])}=\overline{\Lambda
_{t+\tau ,t}^{st}[\rho _{\{y\}}]},  \label{NoMemoryHst}
\end{equation}%
where%
\begin{equation}
\rho _{\{y,x\}}\equiv \rho _{y}\frac{\mathrm{Tr}_{s}(E_{y}\Lambda
_{t,0}^{st}[\rho _{x}])}{\mathrm{Tr}_{s}\overline{(E_{y}\Lambda
_{t,0}^{st}[\rho _{x}])}},\ \ \ \ \ \rho _{\{y\}}\equiv \rho _{y}\frac{%
\mathrm{Tr}_{s}(E_{y}\Lambda _{t,0}^{st}[\rho _{0}])}{\mathrm{Tr}_{s}%
\overline{(E_{y}\Lambda _{t,0}^{st}[\rho _{0}])}}.
\end{equation}%
The left hand side of Eq.~(\ref{NoMemoryHst}) can be read as the system
propagator conditioned to the occurrence of $x$- and $y$-outcomes. The right
hand side implies that this average is independent of the $x$-outcomes,
which in turn implies Markovianity in propagator, Eq.~(\ref{PropaMarkov}).
Notice that white noises are always consistent with Markovianity.

\subsection{Completely positive system-environment dynamics.}

Here the underlying dynamics is defined by a CP bipartite propagator $%
\mathcal{G}_{t,t^{\prime }}.$ Similarly to the previous case, from $%
P_{3}(z,y,x)$ [see Eq.~(\ref{TresCP})] it is possible to write%
\begin{equation}
P_{3}(z|y,x)=\mathrm{Tr}_{se}(E_{z}\mathcal{G}_{t+\tau ,t}[\rho _{y}\otimes
\sigma _{t|y,x}]),  \label{P3CondMicro}
\end{equation}%
where the environment state $\sigma _{t|y,x}$ is%
\begin{equation}
\sigma _{t|y,x}=\frac{\mathrm{Tr}_{s}(E_{y}\mathcal{G}_{t,0}[\rho
_{x}\otimes \sigma _{0}])}{\mathrm{Tr}_{se}(E_{y}\mathcal{G}_{t,0}[\rho
_{x}\otimes \sigma _{0}])}.  \label{RhoBathyx}
\end{equation}%
On the other hand, from $P_{2}(z|y)=P_{2}(z,y)/P_{1}(y)$ it follows that%
\begin{equation}
P_{2}(z|y)=\mathrm{Tr}_{se}(E_{z}\mathcal{G}_{t+\tau ,t}[\rho _{y}\otimes
\sigma _{t|y}]),
\end{equation}%
where here the environment state is%
\begin{equation}
\sigma _{t|y}=\frac{\mathrm{Tr}_{s}(E_{y}\mathcal{G}_{t,0}[\rho _{0}\otimes
\sigma _{0}])}{\mathrm{Tr}_{se}(E_{y}\mathcal{G}_{t,0}[\rho _{0}\otimes
\sigma _{0}])}.
\end{equation}%
Using the arbitrariness of the $z$-measurement, the condition Eq.~(\ref%
{PMarkovian}) leads to%
\begin{eqnarray}
\mathrm{Tr}_{e}(\mathcal{G}_{t+\tau ,t}[\rho _{y}\otimes \sigma _{t|y,x}])
&=&\mathrm{Tr}_{e}(\mathcal{G}_{t+\tau ,t}[\rho _{y}\otimes \sigma _{t|y}]),
\notag \\
\Lambda _{t+\tau ,t}[\rho _{\{y,x\}}] &=&\Lambda _{t+\tau ,t}[\rho _{\{y\}}].
\label{NoMemoryCP}
\end{eqnarray}%
The last equality explicitly shows the validity of Markovianity in
propagator, Eq.~(\ref{PropaMarkov}). The results expressed by Eqs.~(\ref%
{NoMemoryHst}) and~(\ref{NoMemoryCP}) can be extended straightforwardly to
the case in which $n$-measurement processes are performed at successive
times.

\section{Conditions for the validity of DNI\label{DNIConditions}}

Here, we find the conditions under which non-Markovian dynamics could
satisfy DNI. Given the joint probability $P_{2}(z,x)$ and the marginal one $%
P_{3}(z,x)=\sum_{y}P_{3}(z,y,x),$ where the intermediate $y$-measurement is
performed in the basis where the system density matrix is diagonal, DNI is
expressed by the condition%
\begin{equation}
P_{3}(z,x)\overset{DNI}{=}P_{2}(z,x).  \label{DNI_Supp}
\end{equation}%
This equality must be valid independently of which $z$- and $x$-measurement
processes are performed. The case of stochastic Hamiltonian and CP $s$-$e$
dynamics are worked out in a separate way.

\subsection{Stochastic Hamiltonian models}

For this modeling, the joint probability for the three measurement outcomes
reads%
\begin{equation}
\frac{P_{3}(z,y,x)}{P_{1}(x)}=\mathrm{Tr}_{s}\overline{(E_{z}\Lambda
_{t+\tau ,t}^{st}[\rho _{y}])\ \mathrm{Tr}_{s}(E_{y}\Lambda _{t,0}^{st}[\rho
_{x}])}.  \label{P3}
\end{equation}%
Given a $x$-outcome, the (average) system density matrix at time $t,$
denoted as $\rho _{t|x},$\ reads%
\begin{equation}
\rho _{t|x}=\overline{\Lambda _{t,0}^{st}}[\rho _{x}]=\sum_{c}|c_{t}\rangle
\langle c_{t}|p_{t}^{c}.  \label{DiagoNoise}
\end{equation}%
Here, the (in general, time-dependent) projectors $\{\Pi _{t}^{c}\}\equiv
\{|c_{t}\rangle \langle c_{t}|\}$ define the basis where $\rho _{t|x}$ is a
diagonal matrix at time $t.$ The corresponding eigenvalues are the
probabilities $\{p_{t}^{c}\}.$ Notice that in general the projectors $\{\Pi
_{t}^{c}\}$ (and the eigenvalues $\{p_{t}^{c}\})$ depend on which specific $%
x $-measurement process is performed (e.g. direction) and it kind, selective
or non-selective [in which case in Eq.~(\ref{DiagoNoise}) $\rho
_{x}\rightarrow \sum_{x}P_{1}(x)\rho _{x}].$ For simplicity these
dependences are not written explicitly in $\{\Pi _{t}^{c}\}$ neither $%
\{p_{t}^{c}\}.$

The probability $P_{3}(z,x)=\sum_{y}P_{3}(z,y,x),$ by performing the
projective $y$-measurement in the basis $\{\Pi _{t}^{c}\},$ from Eq.~(\ref%
{P3}) reads%
\begin{equation}
\frac{P_{3}(z,x)}{P_{1}(x)}=\sum_{c}\mathrm{Tr}_{s}\overline{(E_{z}\Lambda
_{t+\tau ,t}^{st}[\Pi _{t}^{c}])\ \mathrm{Tr}_{s}(\Pi _{t}^{c}\Lambda
_{t,0}^{st}[\rho _{x}])}.  \label{P3ZXJoint}
\end{equation}%
On the other hand, when performing only two measurements at times $t=0$ and $%
t+\tau ,$ the outcome probability is%
\begin{equation}
\frac{P_{2}(z,x)}{P_{1}(x)}=\mathrm{Tr}_{s}\overline{(E_{z}\Lambda _{t+\tau
,0}^{st}[\rho _{x}])}=\mathrm{Tr}_{s}(E_{z}\Lambda _{t+\tau ,0}[\rho _{x}]),
\label{P2ZX}
\end{equation}%
where the system propagator is $\Lambda _{t,0}[\rho ]=\overline{\Lambda
_{t,0}^{st}}[\rho ].$

DNI is expressed by the condition Eq.~(\ref{DNI_Supp}) where the
probabilities $P_{3}(z,x)$ and $P_{2}(z,x)$ are defined respectively by
Eqs.~(\ref{P3ZXJoint}) and (\ref{P2ZX}). Hence, straightforwardly it follows%
\begin{equation}
\mathrm{Tr}_{s}(E_{z}\Lambda _{t+\tau ,0}[\rho _{x}])\!\!=\!\!\sum_{c}\!%
\mathrm{Tr}_{s}\overline{(E_{z}\Lambda _{t+\tau ,t}^{st}[\Pi _{t}^{c}])%
\mathrm{Tr}_{s}(\Pi _{t}^{c}\Lambda _{t,0}^{st}[\rho _{x}])}.  \label{Previa}
\end{equation}%
By maintaining \textit{fixed} the basis where the $x$- and $z$-measurements
are performed, this relation can accidentally be valid for quantum
non-Markovian dynamics. This property was studied in Ref.~\cite{plenio}.
Nevertheless, in the present approach \textit{this equality must be valid
for arbitrary }$x$\textit{- and }$z$\textit{-measurements.} Using this
stronger condition, which implies arbitrary sets $\{E_{z}\}$ and $\{\rho
_{x}\},$ the previous relation [Eq.~(\ref{Previa})] leads to%
\begin{equation}
\Lambda _{t+\tau ,0}[\rho ]\overset{DNI}{=}\sum_{c}\overline{\Lambda
_{t+\tau ,t}^{st}[\Pi _{t}^{c}]\ \mathrm{Tr}_{s}(\Pi _{t}^{c}\Lambda
_{t,0}^{st}[\rho ])}.  \label{DNI_St}
\end{equation}%
Here, $\rho $ denotes an\textit{\ arbitrary input state,} while $\{\Pi
_{t}^{c}\}$ follows from $\rho _{t}=\Lambda _{t,0}[\rho ]=\sum_{c}\Pi
_{t}^{c}p_{t}^{c}.$ The equality~(\ref{DNI_St}) introduces a severe
constraint to be fulfilled by the underlying quantum stochastic dynamics. It
must be valid for arbitrary $\rho .$

Given that the underlying stochastic dynamics is divisible [Eq.~(\ref%
{HEstocastico})] it follows that%
\begin{equation}
\Lambda _{t+\tau ,0}[\rho ]=\overline{\Lambda _{t+\tau ,t}^{st}[\Lambda
_{t,0}^{st}[\rho ]]}.
\end{equation}%
Now, under the replacing $\Lambda _{t,0}^{st}[\rho ]\rightarrow \mathrm{I}%
_{s}\Lambda _{t,0}^{st}[\rho ]\mathrm{I}_{s},$ where the identity matrix is
written as $\mathrm{I}_{s}=\sum_{c}\Pi _{t}^{c}=\sum_{c}|c_{t}\rangle
\langle c_{t}|,$ the previous equation can be rewritten as%
\begin{eqnarray}
\Lambda _{t+\tau ,0}[\rho ] &=&\sum_{c}\overline{\Lambda _{t+\tau
,t}^{st}[\Pi _{t}^{c}]\ \mathrm{Tr}_{s}(\Pi _{t}^{c}\Lambda _{t,0}^{st}[\rho
])}  \notag \\
&&+\sum_{\substack{ c,\tilde{c}  \\ c\neq \tilde{c}}}\overline{\Lambda
_{t+\tau ,t}^{st}[|c_{t}\rangle \langle \tilde{c}_{t}|]\ \langle
c_{t}|\Lambda _{t,0}^{st}[\rho ]|\tilde{c}_{t}\rangle }.
\end{eqnarray}%
By comparing Eq.~(\ref{DNI_St}) with this expression, it follows that DNI
can equivalently be expressed as%
\begin{equation}
0\overset{DNI}{=}\sum_{\substack{ c,\tilde{c}  \\ c\neq \tilde{c}}}\overline{%
\Lambda _{t+\tau ,t}^{st}[|c_{t}\rangle \langle \tilde{c}_{t}|]\ \langle
c_{t}|\Lambda _{t,0}^{st}[\rho ]|\tilde{c}_{t}\rangle }.
\label{ConditionNoise}
\end{equation}%
This equality must be valid for arbitrary initial states $\rho .$ Given that 
$\Lambda _{t,t^{\prime }}^{st}$ depends on each noise realization, $\Lambda
_{t+\tau ,t}^{st}[|c_{t}\rangle \langle \tilde{c}_{t}|]$ neither $\langle
c_{t}|\Lambda _{t,0}^{st}[\rho ]|\tilde{c}_{t}\rangle $ vanish identically.
The unique general property that relates the propagator $\Lambda
_{t,t^{\prime }}^{st}$ and the states $|c_{t}\rangle \langle \tilde{c}_{t}|$
is$\ \overline{\langle c_{t}|\Lambda _{t,0}^{st}[\rho ]|\tilde{c}_{t}\rangle 
}=\langle c_{t}|\Lambda _{t,0}[\rho _{0}]|\tilde{c}_{t}\rangle =\langle
c_{t}|\rho _{t}|\tilde{c}_{t}\rangle =0.$ Thus, \textit{DNI} [Eq.~(\ref%
{ConditionNoise}) with arbitrary initial conditions] \textit{can only be
satisfied when the noise fluctuations are white.}

A more formal demonstration of the previous affirmation is developed below.
Taking an arbitrary initial condition $\rho _{0}=|c_{0}\rangle \langle
c_{0}|,$ when $\tau =0$ Eq.~(\ref{ConditionNoise}) becomes%
\begin{equation}
0=\sum_{\substack{ c,\tilde{c}  \\ c\neq \tilde{c}}}\overline{|c_{t}\rangle
\langle \tilde{c}_{t}|\langle c_{t}|c_{t}^{st}\rangle \langle c_{t}^{st}|%
\tilde{c}_{t}\rangle }=\sum_{\substack{ c,\tilde{c}  \\ c\neq \tilde{c}}}%
|c_{t}\rangle \langle \tilde{c}_{t}|\overline{\langle
c_{t}|c_{t}^{st}\rangle \langle c_{t}^{st}|\tilde{c}_{t}\rangle }.
\label{CerolAlways}
\end{equation}%
Here, we have defined the scalar products%
\begin{equation}
\langle c_{t}|c_{t}^{st}\rangle \equiv \langle
c_{t}|U_{t,0}^{st}|c_{0}\rangle ,\ \ \ \ \ \ \langle c_{t}^{st}|\tilde{c}%
_{t}\rangle \equiv \langle c_{0}|U_{t,0}^{\dag st}|\tilde{c}_{t}\rangle ,
\end{equation}%
where the wave vector propagator $U_{t,0}^{st}$ is defined from $\Lambda
_{t,0}^{st}[\rho ]=U_{t,0}^{st}\rho U_{t,0}^{\dag st}.$ The equality~(\ref%
{CerolAlways}) is always satisfied because $\overline{\langle
c_{t}|c_{t}^{st}\rangle \langle c_{t}^{st}|\tilde{c}_{t}\rangle ,}=\langle
c_{t}|\rho _{t}|\tilde{c}_{t}\rangle =0$ [see Eq.~(\ref{DiagoNoise})]. At
time $t+\tau ,$ Eq.~(\ref{ConditionNoise}) can be rewritten as%
\begin{equation}
0\overset{DNI}{=}\sum_{\substack{ c,\tilde{c}  \\ c\neq \tilde{c}}}\overline{%
|c_{t+\tau |t}^{st}\rangle \langle \tilde{c}_{t+\tau |t}^{st}|\ \langle
c_{t}|c_{t}^{st}\rangle \langle c_{t}^{st}|\tilde{c}_{t}\rangle },
\label{CambioBase}
\end{equation}%
where we have introduced the stochastic states%
\begin{equation}
|c_{t+\tau |t}^{st}\rangle \equiv U_{t+\tau ,t}^{st}|c_{t}\rangle ,\ \ \ \ \
\ \ \ \langle \tilde{c}_{t+\tau |t}^{st}|\equiv \langle \tilde{c}%
_{t}|U_{t+\tau ,t}^{\dag st}.
\end{equation}%
\textit{Fixing} a noise realization up to time $t,$ the operator $|c_{t+\tau
|t}^{st}\rangle \langle \tilde{c}_{t+\tau |t}^{st}|$ [in the interval $%
(t,t+\tau )]$ has the same properties as $|c_{t}\rangle \langle \tilde{c}%
_{t}|$ in Eq.~(\ref{CerolAlways}). In fact, $\langle \tilde{c}_{t+\tau
|t}^{st}|c_{t+\tau |t}^{st}\rangle =0$ because $U_{t+\tau ,t}^{st}$ is a
(stochastic) unitary operator. Therefore, the (conditional) average in the
interval $(t,t+\tau )$ is not null. Consequently, the same condition (a
non-vanishing average) is inherited after taking arbitrary realizations in
the interval $(0,t),$ which contradicts the equality~(\ref{CambioBase}).
Similarly to Eq.~(\ref{CerolAlways}), this contradiction is raised up (for $%
\tau >0)$ under the condition%
\begin{equation}
0\overset{DNI}{=}\sum_{\substack{ c,\tilde{c}  \\ c\neq \tilde{c}}}\overline{%
|c_{t+\tau |t}^{st}\rangle \langle \tilde{c}_{t+\tau |t}^{st}|}\ \overline{%
\langle c_{t}|c_{t}^{st}\rangle \langle c_{t}^{st}|\tilde{c}_{t}\rangle },
\end{equation}%
which correspond to a white noise limit. Consequently, for stochastic
Hamiltonian dynamics we conclude that \textit{DNI is valid if and only if
the dynamics is Markovian}. Equivalently, if for some measurement processes
the dynamic is non-Markovian it always violate DNI for some $x$- and $z$%
-measurement processes.

\subsection{Completely positive system-environment dynamics}

In this case, the joint probability for the three measurement outcomes reads%
\begin{equation}
\frac{P_{3}(z,y,x)}{P_{1}(x)}=\mathrm{Tr}_{se}(E_{z}\mathcal{G}_{t+\tau
,t}[\rho _{y}\otimes \mathrm{Tr}_{s}(E_{y}\mathcal{G}_{t,0}[\rho _{x}\otimes
\sigma _{0}])]).  \label{3CP}
\end{equation}%
Given an $x$-outcome, the system density matrix at time $t$\ reads%
\begin{equation}
\rho _{t|x}\equiv \mathrm{Tr}_{e}(\mathcal{G}_{t,0}[\rho _{x}\otimes \sigma
_{0}])=\sum_{c}|c_{t}\rangle \langle c_{t}|p_{t}^{c}.
\end{equation}%
By performing the $y$-measurement in the basis $\{\Pi
_{t}^{c}\}=\{|c_{t}\rangle \langle c_{t}|\}$\ the marginal probability $%
P_{3}(z,x),$ from Eq.~(\ref{3CP}), reads%
\begin{equation}
\frac{P_{3}(z,x)}{P_{1}(x)}=\sum_{c}\mathrm{Tr}_{se}(E_{z}\mathcal{G}%
_{t+\tau ,t}[\Pi _{t}^{c}\otimes \mathrm{Tr}_{s}(\Pi _{t}^{c}\mathcal{G}%
_{t,0}[\rho _{x}\otimes \sigma _{0}])]).  \label{Q3ZX}
\end{equation}%
When performing only two measurements at times $t=0$ and $t+\tau ,$ the
outcome joint probability is%
\begin{equation}
\frac{P_{2}(z,x)}{P_{1}(x)}=\mathrm{Tr}_{se}(E_{z}\mathcal{G}_{t+\tau
,0}[\rho _{x}\otimes \sigma _{0}])=\mathrm{Tr}_{s}(E_{z}\Lambda _{t+\tau
,0}[\rho _{x}]).  \label{Q2ZX}
\end{equation}%
Here, the (CP) system density matrix propagator is $\Lambda _{t,0}[\rho ]=%
\mathrm{Tr}_{e}(\mathcal{G}_{t,0}[\rho \otimes \sigma _{0}]),$ where $\sigma
_{0}$ is the initial environment state.

The validity of DNI [Eq.~(\ref{DNI_Supp})], from Eqs.~(\ref{Q3ZX}) and (\ref%
{Q2ZX}), implies that%
\begin{eqnarray}
&&\mathrm{Tr}_{s}(E_{z}\Lambda _{t+\tau ,0}[\rho _{x}]) \\
&=&\sum_{c}\mathrm{Tr}_{se}(E_{z}\mathcal{G}_{t+\tau ,t}[\Pi _{t}^{c}\otimes 
\mathrm{Tr}_{s}(\Pi _{t}^{c}\mathcal{G}_{t,0}[\rho _{x}\otimes \sigma
_{0}])]).  \notag
\end{eqnarray}%
As in the previous stochastic Hamiltonian case, maintaining \textit{fixed}
the basis where the $x$- and $z$-measurements are performed, this relation
can be accidentally valid~\cite{plenio}. Nevertheless, given that in the
present approach the sets $\{E_{z}\}$ and $\{\rho _{x}\}$ are arbitrary
ones, it follows the condition%
\begin{equation}
\Lambda _{t+\tau ,0}[\rho ]\overset{DNI}{=}\sum_{c}\mathrm{Tr}_{e}(\mathcal{G%
}_{t+\tau ,t}[\Pi _{t}^{c}\otimes \mathrm{Tr}_{s}(\Pi _{t}^{c}\mathcal{G}%
_{t,0}[\rho \otimes \sigma _{0}])]).  \label{DNI_CCPP}
\end{equation}%
Here, $\rho $ denotes an arbitrary initial system state, while $\{\Pi
_{t}^{c}\}$ follows from $\rho _{t}=\Lambda _{t,0}[\rho ]=\sum_{c}\Pi
_{t}^{c}p_{t}^{c}.$ On the other hand, this equality must be valid for
arbitrary initial system states $\rho .$

Given that the underlying \textit{bipartite} propagator is divisible [Eq.~(%
\ref{Gbipartito})] it follows that%
\begin{equation}
\Lambda _{t+\tau ,0}[\rho ]=\mathrm{Tr}_{e}(\mathcal{G}_{t+\tau ,t}[\mathcal{%
G}_{t,0}[\rho \otimes \sigma _{0}]]).
\end{equation}%
Under the replacing $\mathcal{G}_{t,0}[\rho \otimes \sigma _{0}]\rightarrow 
\mathrm{I}_{s}\mathcal{G}_{t,0}[\rho \otimes \sigma _{0}]\mathrm{I}_{s},$
where the identity matrix is written as $\mathrm{I}_{s}=\sum_{c}\Pi
_{t}^{c}=\sum_{c}|c_{t}\rangle \langle c_{t}|,$ the previous equation leads
to%
\begin{equation}
\begin{array}{c}
\Lambda _{t+\tau ,0}[\rho ]=\sum_{c}\mathrm{Tr}_{e}(\mathcal{G}_{t+\tau
,t}[\Pi _{t}^{c}\otimes \mathrm{Tr}_{s}(\Pi _{t}^{c}\mathcal{G}_{t,0}[\rho
\otimes \sigma _{0}])]) \\ 
\\ 
\ \ \ \ \ \ \ \ \ \ +\sum_{\substack{ c,\tilde{c}  \\ c\neq \tilde{c}}}%
\mathrm{Tr}_{e}(\mathcal{G}_{t+\tau ,t}[\Pi _{t}^{c\tilde{c}}\otimes \mathrm{%
Tr}_{s}(\Pi _{t}^{\tilde{c}c}\mathcal{G}_{t,0}[\rho \otimes \sigma _{0}])]),%
\end{array}
\label{PropaConlaBasemovil}
\end{equation}%
where $\Pi _{t}^{c\tilde{c}}=|c_{t}\rangle \langle \tilde{c}_{t}|.$ By
comparing this expression with Eq.~(\ref{DNI_CCPP}) it follows that DNI can
equivalently be expressed as%
\begin{equation}
0\overset{DNI}{=}\sum_{\substack{ c,\tilde{c}  \\ c\neq \tilde{c}}}\mathrm{Tr%
}_{e}(\mathcal{G}_{t+\tau ,t}[\Pi _{t}^{c\tilde{c}}\otimes \mathrm{Tr}%
_{s}(\Pi _{t}^{\tilde{c}c}\mathcal{G}_{t,0}[\rho \otimes \sigma _{0}])]),
\label{THECondition}
\end{equation}%
where $\rho $ is an arbitrary system state.

\subsubsection{Markovian solution}

The DNI condition Eq. (\ref{THECondition}) is always satisfied if the
bipartite propagator satisfies%
\begin{equation}
\mathrm{Tr}_{e}(\mathcal{G}_{t+\tau ,t}[\rho \otimes \sigma ])=\Lambda
_{t+\tau ,t}[\rho ]\mathrm{Tr}_{e}(\sigma ).  \label{MarkovBipartito}
\end{equation}%
In fact, by definition of the projectors $\Pi _{t}^{\tilde{c}c},$ the
equality%
\begin{equation}
\mathrm{Tr}_{se}(\Pi _{t}^{\tilde{c}c}\mathcal{G}_{t,0}[\rho \otimes \sigma
_{0}])=0,
\end{equation}%
is always satisfied. Eq.~(\ref{MarkovBipartito}) expresses that the system
propagator does not depends on the previous system history. Thus, it
guarantees the Markovianity of the reduced system dynamics.

\subsubsection{Superclassical non-Markovian dynamics}

In contrast to the stochastic Hamiltonian case, here it is possible to
define a specific class of quantum non-Markovian dynamics that fulfill DNI.
They are termed as superclassical dynamics~\cite{super}. These kind of
evolutions only emerges when considering a non-unitary (Lindblad) $s$-$e$
dynamics. They are briefly described below.

\textit{Solutions with vanishing quantum discord}: Eq.~(\ref{THECondition})
is identically satisfied if one assume that%
\begin{equation}
0=\mathrm{Tr}_{s}(\Pi _{t}^{\tilde{c}c}\mathcal{G}_{t,0}[\rho \otimes \sigma
_{0}]),\ \ \ \ \ c\neq \tilde{c}.  \label{DiscoCondition}
\end{equation}%
This condition implies that, at an arbitrary time $t,$ the bipartite
propagator can be written as%
\begin{equation}
\mathcal{G}_{t,0}[\rho \otimes \sigma _{0}]=\sum_{c}|c_{t}\rangle \langle
c_{t}|\otimes \sigma _{t}^{(c)}=\sum_{c}\Pi _{t}^{c}\otimes \sigma
_{t}^{(c)},  \label{Discordia}
\end{equation}%
where $\{\sigma _{t}^{(c)}\}$ are environment states. Thus, in this case the
bipartite evolution, \textit{independently of the system initial condition},
does not generate any quantum discord.

\textit{Solutions with non-vanishing quantum discord: }When the condition~(%
\ref{DiscoCondition}) is not fulfilled [equivalently Eq.~(\ref{Discordia})
is not fulfilled] the bipartite evolution generates quantum discord. Even in
this case it is possible to satisfy the DNI condition~(\ref{THECondition}).
When this situation occurs, the non-Markovian system dynamics induced by the
reservoir is always equivalent to the system dynamics induced in absence of
quantum discord generation. This conclusion follows straightforwardly from
Eqs.~(\ref{PropaConlaBasemovil}) and~(\ref{THECondition}). This property
expresses that, when DNI is fulfilled, the $s$-$e$ correlations that lead to
discord generation do not affect the system propagator.

\subsubsection{Unitary system-environment models are incompatible with
superclassicality}

Here we show that superclassicality does not arises when the underlying $s$-$%
e$ dynamics is unitary, that is, when%
\begin{equation}
\rho _{t}^{se}=\mathcal{G}_{t,0}[\rho _{0}\otimes \sigma
_{0}]=U_{t,0}^{se}(\rho _{0}\otimes \sigma _{0})U_{t,0}^{\dagger se},
\label{PropaUnil}
\end{equation}%
where $U_{t,t_{0}}^{se}\equiv \exp [-i\int_{t_{0}}^{t}dt^{\prime
}H_{T}(t^{\prime })].$ Contrarily to dissipative models, unitary $s$-$e$\
dynamics \textit{always} leads to quantum discord generation, neither
fulfills the DNI condition Eq.~(\ref{THECondition}). These results are
demonstrated below.

The condition of null discord generation, Eq.~(\ref{DiscoCondition}), here
can be written as%
\begin{equation}
0=\langle c_{t}|U_{t,0}^{se}(\rho \otimes \sigma _{0})U_{t,0}^{\dagger se}|%
\tilde{c}_{t}\rangle ,\ \ \ \ \ c\neq \tilde{c}.  \label{DicoUnitario}
\end{equation}%
On the other hand, the basis of states $\{|c_{t}\rangle \}$ and $%
\{|c_{0}\rangle \},$ where the system state is a diagonal matrix at an
arbitrary time $t$ and at $t=0$ respectively, are always related by a
unitary transformation,%
\begin{equation}
|c_{t}\rangle =\exp [-i\mathbb{H}_{t}]|c_{0}\rangle .
\end{equation}%
In general, the Hermitian operator $\mathbb{H}_{t}$ depends on both the
total Hamiltonian and the $s$-$e$ initial state. Nevertheless, it does not
depend on the states $\{|c_{0}\rangle \}.$ Taking an arbitrary system
initial condition with the structure $\rho _{0}=|\bar{c}_{0}\rangle \langle 
\bar{c}_{0}|,$ where $|\bar{c}_{0}\rangle \in \{|c_{0}\rangle \},$ the null
quantum discord condition Eq.~(\ref{DicoUnitario}) can be rewritten as%
\begin{equation}
0=\langle c_{0}|e^{+i\mathbb{H}_{t}}U_{t,0}^{se}|\bar{c}_{0}\rangle \sigma
_{0}\langle \bar{c}_{0}|U_{t,0}^{\dagger se}e^{-i\mathbb{H}_{t}}|\tilde{c}%
_{0}\rangle .
\end{equation}%
This equation, which must be valid for arbitrary basis $\{|c_{0}\rangle \},$
is \textit{inconsistent} if the propagator $U_{t,0}^{se}$ intrinsically
couples the system and the environment. In fact, the equality $\langle
c_{0}|e^{-i\mathbb{H}_{t}}U_{t,0}^{se}|c_{0}^{\prime }\rangle =0$ is
consistently satisfied only when the total Hamiltonian $%
H_{T}=H_{s}+H_{e}+H_{se}$ does not introduce any $s$-$e$ coupling, $%
H_{se}=0. $

The unavoidable discord generation induced by unitary $s$-$e$ interactions
also \textit{forbids} the fulfilment of the DNI condition Eq.~(\ref%
{THECondition}). In fact, Eqs.~(\ref{PropaConlaBasemovil}) and~(\ref%
{THECondition}) imply that when DNI is fulfilled the system propagator in
absence or presence of discord generation is the same, the former propagator
being inexistent. A technical demonstration of this impossibility is
provided below.

Using that $\mathrm{I}_{s}=\sum_{s}|s\rangle \langle s|=\sum_{\tilde{s}}|%
\tilde{s}\rangle \langle \tilde{s}|,$ where $\{|s\rangle \}$ and $\{|\tilde{s%
}\rangle \}$ are arbitrary complete basis in the system Hilbert space, the
DNI condition Eq.~(\ref{THECondition}), with the unitary propagator~(\ref%
{PropaUnil}), assumes the structure%
\begin{equation}
0\overset{DNI}{=}\sum_{s,\tilde{s}}|s\rangle \langle \tilde{s}|\ \sum 
_{\substack{ c,\tilde{c}  \\ c\neq \tilde{c}}}\mathrm{Tr}_{e}(\langle
s|U_{t+\tau ,t}^{se}|c_{t}\rangle \sigma _{t}^{c\tilde{c}}\langle \tilde{c}%
_{t}|U_{t+\tau ,t}^{\dagger se}|\tilde{s}\rangle ),  \label{Cerapio}
\end{equation}%
where $\sigma _{t}^{c\tilde{c}}\equiv \langle c_{t}|U_{t,0}^{se}(\rho
\otimes \sigma _{0})U_{t,0}^{\dagger se}|\tilde{c}_{t}\rangle \neq 0.$
Taking into account that $\mathrm{Tr}_{e}(\sigma _{t}^{c\tilde{c}})=0,$
given the arbitrariness of the basis $\{|s\rangle \}$ and $\{|\tilde{s}%
\rangle \},$ it follows that a sufficient condition for fulfilling Eq.~(\ref%
{Cerapio}) is%
\begin{equation}
\langle \tilde{c}_{t}|U_{t+\tau ,t}^{\dagger se}|\tilde{s}\rangle \langle
s|U_{t+\tau ,t}^{se}|c_{t}\rangle \overset{DNI}{\simeq }\mathrm{I}_{e},\ \ \
\ \ \ \forall (|s\rangle ,|\tilde{s}\rangle ,|c_{t}\rangle ,|\tilde{c}%
_{t}\rangle ).  \label{Inco1}
\end{equation}%
Nevertheless, this proportionality can only be consistently fulfilled when
the system and the environment are not coupled by the interaction
Hamiltonian.

In general, introducing in Eq.~(\ref{Cerapio}) the explicit expression for $%
\sigma _{t}^{c\tilde{c}},$ taking $\rho _{0}=|\bar{c}_{0}\rangle \langle 
\bar{c}_{0}|,$ where $|\bar{c}_{0}\rangle $ is an arbitrary system state,
using the arbitrariness of the basis $\{|s\rangle \}$ and $\{|\tilde{s}%
\rangle \},$ it follows the condition 
\begin{equation}
0\overset{DNI}{=}\mathrm{Tr}_{e}(R_{s,\tilde{s},\bar{c}_{0}}\sigma _{0}),\ \
\ \ \ \ \forall (|s\rangle ,|\tilde{s}\rangle ,|c_{0}\rangle ),
\label{Inco22}
\end{equation}%
where the environment operator $R_{s,\tilde{s},\bar{c}_{0}}=R$ is%
\begin{equation}
R=\sum_{\substack{ c,\tilde{c}  \\ c\neq \tilde{c}}}\langle \bar{c}%
_{0}|U_{t,0}^{\dagger se}|\tilde{c}_{t}\rangle \langle \tilde{c}%
_{t}|U_{t+\tau ,t}^{\dagger se}|\tilde{s}\rangle \langle s|U_{t+\tau
,t}^{se}|c_{t}\rangle \langle c_{t}|U_{t,0}^{se}|\bar{c}_{0}\rangle .
\end{equation}%
Given that each factor corresponds to the bipartite propagator projected on
an arbitrary system state, the DNI condition can only be fulfilled when
there is not any $s$-$e$ coupling.

We conclude that unitary bipartite evolutions are incompatible with
superclassicality. Equivalently, for unitary $s$-$e$ models DNI can only be
satisfied in a Markovian limit. The corresponding condition [Eq.~(\ref%
{MarkovBipartito})] only applies when a Born-Markov approximation~\cite%
{breuerbook}\ applies, even in the interval between measurement processes.


\begin{thebibliography}{99}
\bibitem{demolition} The DNI property of Markovian dynamics is an extension
to time-evolutions of a well-established result: when an observable and the
pre-measurement state commutate, the density matrix remains the same after a
(non-selective) measurement process. This non-invasive characteristic is a
statistical property that is valid for both \textquotedblleft direct
measurement\textquotedblright\ and \textquotedblleft indirect
measurement\textquotedblright\ processes, non-demolition measurements being
included in the last class.
\bibitem{QRT} $\Lambda $-Markovianity implies that the propagator is
divisible and that a quantum regression formula applies. Nevertheless, the
inverse implication is not valid in general, that is, divisibility does not
implies $\Lambda $-Markovianity (see for example second paper in Ref.~\cite%
{budiniCPF} and \cite{QRTVacchini}).
\end{thebibliography}
\end{document}